\documentclass[preprint]{aastex}
\usepackage[english]{babel}
\usepackage{graphicx}
\usepackage{subfigure}
\usepackage{placeins}

\usepackage{array}   
\usepackage{booktabs} 
\usepackage{multirow} 

\usepackage{adjustbox} 

\usepackage{color}  

\usepackage{multicol}
\usepackage{adjustbox} 

\usepackage{natbib}

\newcommand{\aftr}{}

\shorttitle{Time-resolved emission from bright hot pixels }
\shortauthors{Djorgovski et al.}

\begin{document}


\title{Time-resolved emission from bright hot pixels  of an active region 
    observed in the EUV band with SDO/AIA and 
    multi-stranded loop modeling}


\author{E. Tajfirouze\altaffilmark{1}, F.Reale\altaffilmark{1,2}, A. Petralia\altaffilmark{1}, P. Testa\altaffilmark{3}}
\affil{\altaffilmark{1}Dipartimento di Fisica e Chimica, Universit\`a di Palermo, Piazza del Parlamento 1, 90134, Italy}
\affil{\altaffilmark{2}INAF-Osservatorio Astronomico di Palermo ``G.S. Vaiana'', Piazza del Parlamento 1, 90134, Italy}
\affil{\altaffilmark{3}Harvard-Smithsonian Center for Astrophysics, 60 Garden Street, Cambridge, MA 02138, USA}
\email{aastex-help@aas.org}



\begin{abstract}

Evidence for small amounts of very hot plasma has been found in active regions and might be the indication of an impulsive heating, released at spatial scales smaller than the cross section of a single loop. We investigate the heating and substructure of coronal loops in the core of one such active region by analyzing the light curves in the smallest resolution elements of solar observations in two EUV channels (94~\AA\ and 335~\AA) from the Atmospheric Imaging Assembly on-board the Solar Dynamics Observatory. We model the evolution of a bundle of strands heated by a storm of nanoflares by means of a hydrodynamic 0D loop model (EBTEL). 
The light curves obtained from the random combination of those of single strands are compared to the observed light curves either in a single pixel or in a row of pixels, simultaneously in the two channels and using two independent methods:  an artificial intelligent system (Probabilistic Neural Network, PNN) and a simple cross-correlation technique. 
We explore the space of the parameters to constrain the distribution of the heat pulses, their duration and their spatial size, and, as a feedback on the data, their signatures on the light curves. 
From both methods the best agreement is obtained for a relatively large population of events (1000) with a short duration (less than 1 min) and a relatively shallow distribution (power law with index 1.5) in a limited energy range (1.5 decades).  The feedback on the data indicates that bumps in the light curves, especially in the 94~\AA\ channel, are signatures of a heating excess that occurred a few minutes before.

\end{abstract}


\section{Introduction}

The bright curved magnetic flux tubes called coronal loops are the building blocks of the confined solar corona. According to one popular scenario, coronal loops are mainly heated by short and intense energy pulses, called nanoflares (\citealt{Par88}, \citealt{Car94}, \citealt{Rea14}). These pulses might heat the plasma temporarily to temperatures well above the average coronal temperature, and therefore, this scenario is supported by the evidence for small amounts of ultra-hot ($> 5$ MK) plasma in non-flaring active regions (\citealt{McT09}, \citealt{Rea09}, \citealt{Re09}, \citealt{Tes11}, \citealt{Mic12}, \citealt{Te12}, \citealt{Bro14}, \citealt{Pet14}, \citealt{Cas15}).

According to several recent works (\citealt{Kli08}, \citealt{car12},  \citealt{Via12}),  heat pulses are released at spatial scales smaller than the typical resolved loop cross section. A single loop must then be sub-structured into a bundle of thin strands where the plasma can move and transport energy along the magnetic field lines independently of the others, under the effect of each localised heat pulse. Since the strands are under-resolved at the moment, the cross-section of the smallest components is under debate (\citealt{Pet13}, \citealt{Bro12}). On the other hand, the very efficient thermal conduction along the magnetic field lines at coronal temperatures inhibits also the measurement of the duration of the individual heat pulse.

This work extends the analysis of an active region that has shown evidence of small amounts of very hot plasma ($> 5$ MK) in most of the region core and out of proper flares (\citealt{Rea09}, \citealt{Rea11}, \citealt{Te12}). This area is appropriate to search for signatures of small scale heat pulses.

We investigate the emission variability at the smallest possible scale and look for signatures of heating and possibly elementary heating events.
We model the Extreme Ultraviolet emission observed by the Solar Dynamic Observatory (SDO) from the active region core using a hydrodynamic model of bundles of strands heated impulsively. In the active region core we extract the light curves of both a sample single pixel and a row of neighboring pixels in different channel bands. 
We generate model light curves by summing over random events \aftr{with different energy} and we compare them to observed ones using two independent methods: an artificial intelligent system based on a Probabilistic Neural Network (PNN) and a simple cross-correlation technique.
The simultaneous comparison of the light curves in different channels allows us to address the multi-temperature structure of the loop that is a key point when we consider such a structured heating.

\begin{figure*}[!t]
\centering

\setlength\fboxsep{0pt}
\setlength\fboxrule{2pt}
\includegraphics[width=16cm]{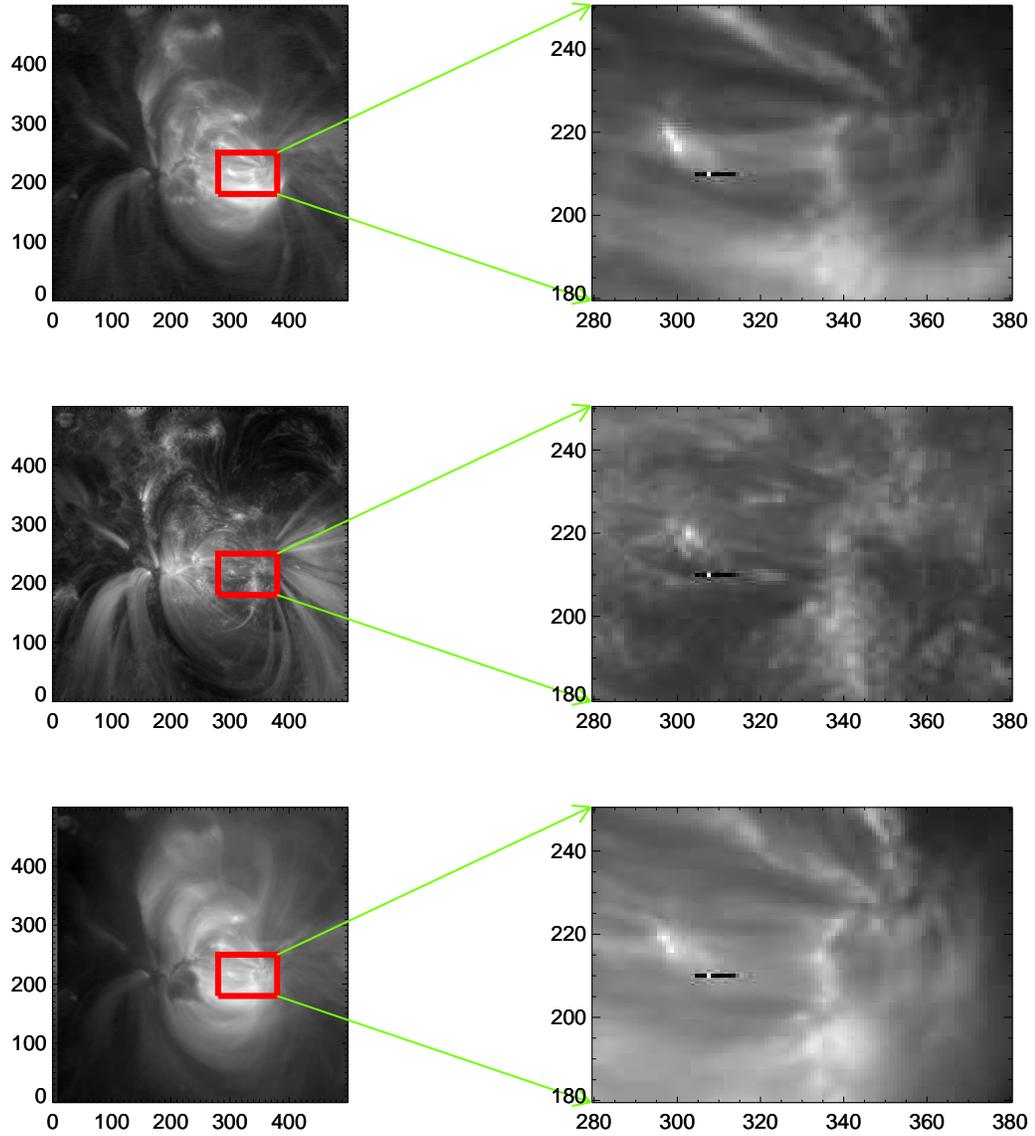}
\caption{Left column: from up to down is the images of the active region in the 94~\AA, 171~\AA\ and 335~\AA\ channels respectively. Right column: zoomed view of the insets in the left images. The single pixel (white) and row of pixels (black) for analysis are marked. } 

\label{fig1} 
\end{figure*}

\begin{figure*}[!t]
\centering

\setlength\fboxsep{0pt}
\setlength\fboxrule{2pt}
\includegraphics[width=16.cm]{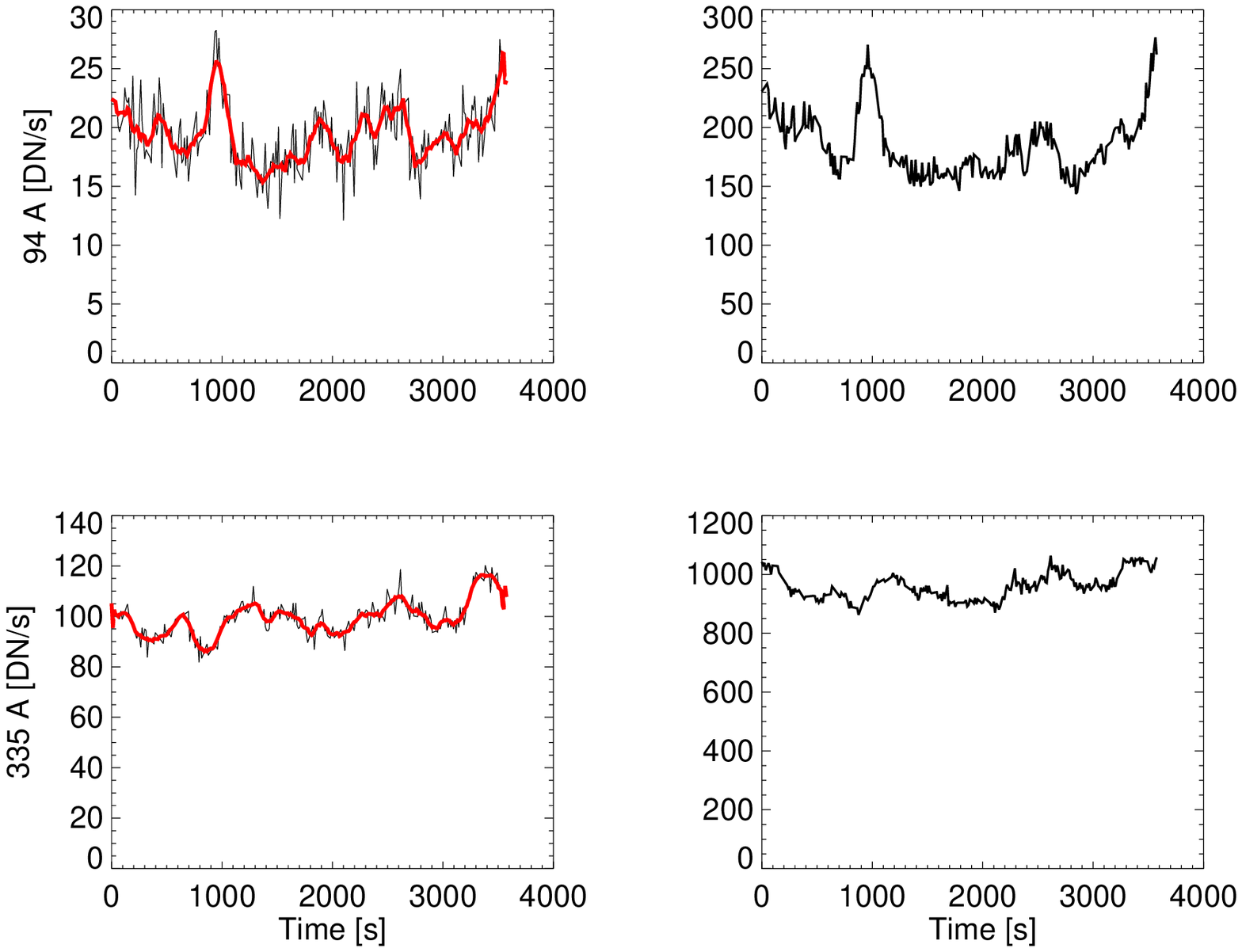}
\caption{Light curves in two AIA channels (up to down: 94~\AA\ and 335~\AA) from the single pixel (left column) and the row of pixels (right column) indicated in figure~\ref{fig1}. \aftr{Smoothed light curves of the single pixel are also shown (red).}} 
\label{fig2} 
\end{figure*}

\section{Observations}
\label{sec:obs}

We study an active region with evidence for small quantities of very hot plasma that may be a signature of short and intense heat pulses \citep{Rea11}. The data and their basic processing have been already described in \cite{Rea11}. We select some specific pixels in the core of this active region. 
In particular, we initially choose the observations in three EUV channels, namely the 94~\AA, 171~\AA\ and 335~\AA\ that are \textbf{most} sensitive to the emission of plasma in a broad range of temperatures,  i.e. at 6 MK, 1 MK and 3 MK respectively. The response function in the 94~\AA\ channel is double peaked, with a cooler peak below ~1MK (\citealt{Boe12}, \citealt{Boe14}, \citealt{Tes12}) . However, in active region cores the hotter peak is generally dominant (e.g., \citealt{Te12}). 

For our study we need the highest possible space and time resolution to try to capture the small temporal and spatial scales expected for nano-flare heating. Therefore, we consider the smallest scale pixel region, i.e. single pixels, and the full time resolution. Figure~\ref{fig1} shows the active region core in the 3 AIA channels and the  location of the pixels where we have extracted light curves. We skipped the 171~\AA\ channel for further analysis, because it is affected  severely by underlying moss emission, i.e., emission coming from the bright footpoints of hot loops (\citealt{Per94}, \citealt{Fle99}). By definition, this emission is not included in our modeling, which comprises only the coronal part of the loops. Since this emission is  out of the scope of our work, from now on we will address only the comparison with the 94 \AA\ and 335 \AA\ channels, which instead are sensitive to the body of the coronal loops. \aftr{We have ascertained that the low temperature contribution to the 94~\AA\ channel is negligible (by rescaling the emission in the 171~\AA\ channel, e.g., \citealt{Rea11}).}

We consider 246 and 228 successive images in the two respective channels,  corresponding to timeseries of total duration of $\sim 1$ hour, i.e., much longer than typical loop plasma cooling times \citep{Rea14}. \aftr{The exposure time is 2.9 s in both channels.} We carefully aligned images between the two different channels via cross correlation using the align\_cube\_correl procedure in SSW. The time distance between the images is typically 12 seconds, but there are some gaps up to $\sim$ 1 minute. We select a single pixel in the bright and filamented core observed in the 94~\AA\ channel, where there is evidence of very hot plasma \citep{Rea11}. It is one where the light curve does not show prominent single-time spikes, which typically affect other pixels. We have also tried to improve for signal-to-noise ratio. Our choice has been to sum the emission in a few nearby pixels. The orientation of observation frame vs the active region shows that many structures are practically aligned in a left-right direction. Actually, we have identified a few rows of pixels where the emission shows a coherent time behaviour, i.e. they probably intercept the same loop strands where the confined plasma evolves coherently. We select a row of 9 pixels which includes the single pixel.
Figure~\ref{fig2} shows the light curves for the single pixel and for the row of 9 pixels. \aftr{To better show the trends and features of the single pixel emission we also show the light curves smoothed with a boxcar of 8 data points.}
The single pixel allows for the maximum possible sensitivity to emission variations, but is affected by significant noise from limited photon statistics. The row of pixels reduces the photon noise but also the sensitivity to variations.
The light curves show different amplitude of fluctuations in the two channels. The evolution in the 335 \AA\ channel is rather smooth, with an overall variation range of  $\sim 30$\%. In the 94 \AA\ channel we see larger fluctuations and even a localised peak $\sim 50$\% above the average with a duration of $\sim 5 \,min$.
The bump shows equal rise and decay times. We find similar trends and features and similar results of our analysis also for other pixels.

From the observation we estimate that the loops inside the active region core have a length of  $\approx 5 \times 10^{9} \,cm$  and we will assume this as our reference loop length from now on.

\section{The analysis}

\subsection{The Loop Model} \label{sec:ebtel}

Our aim is to analyse the time variation of the loop emission. In the reasonable assumption that the plasma evolution does not change much from one position to the other inside a single strand that composes the loop, we \aftr{focus on} the description of the strand population, of the parameters of the related storm of nano-flares, and of how they \aftr{combine} to produce the total observed emission. 
In this scenario, the coronal average quantities in each strand provide enough information to describe the strand properties, but we need a good description of their time evolution.

\begin{figure}[!ht]               
\centering
 \subfigure[]{\includegraphics[width=8cm]{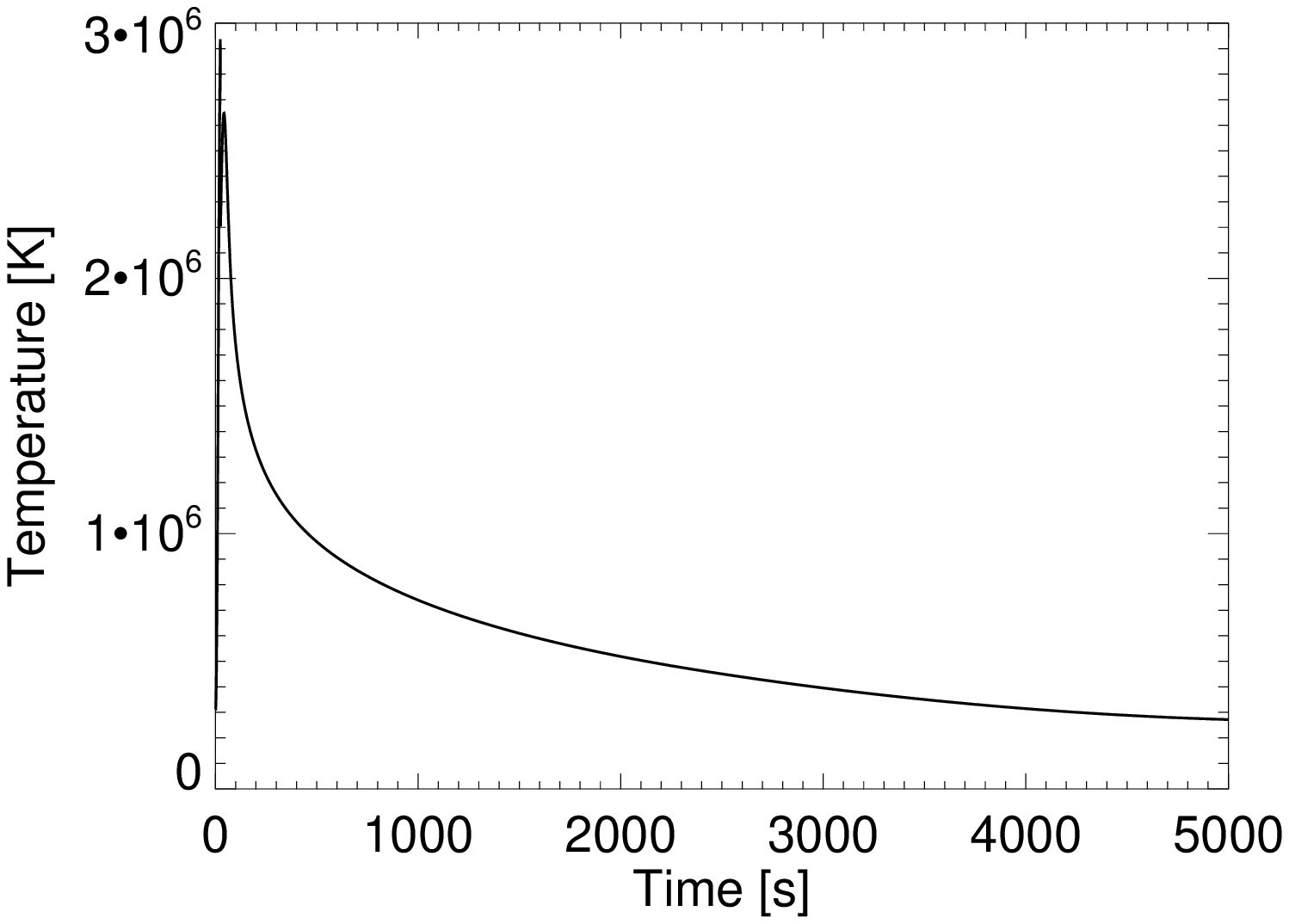}}
 \subfigure[]{\includegraphics[width=8cm]{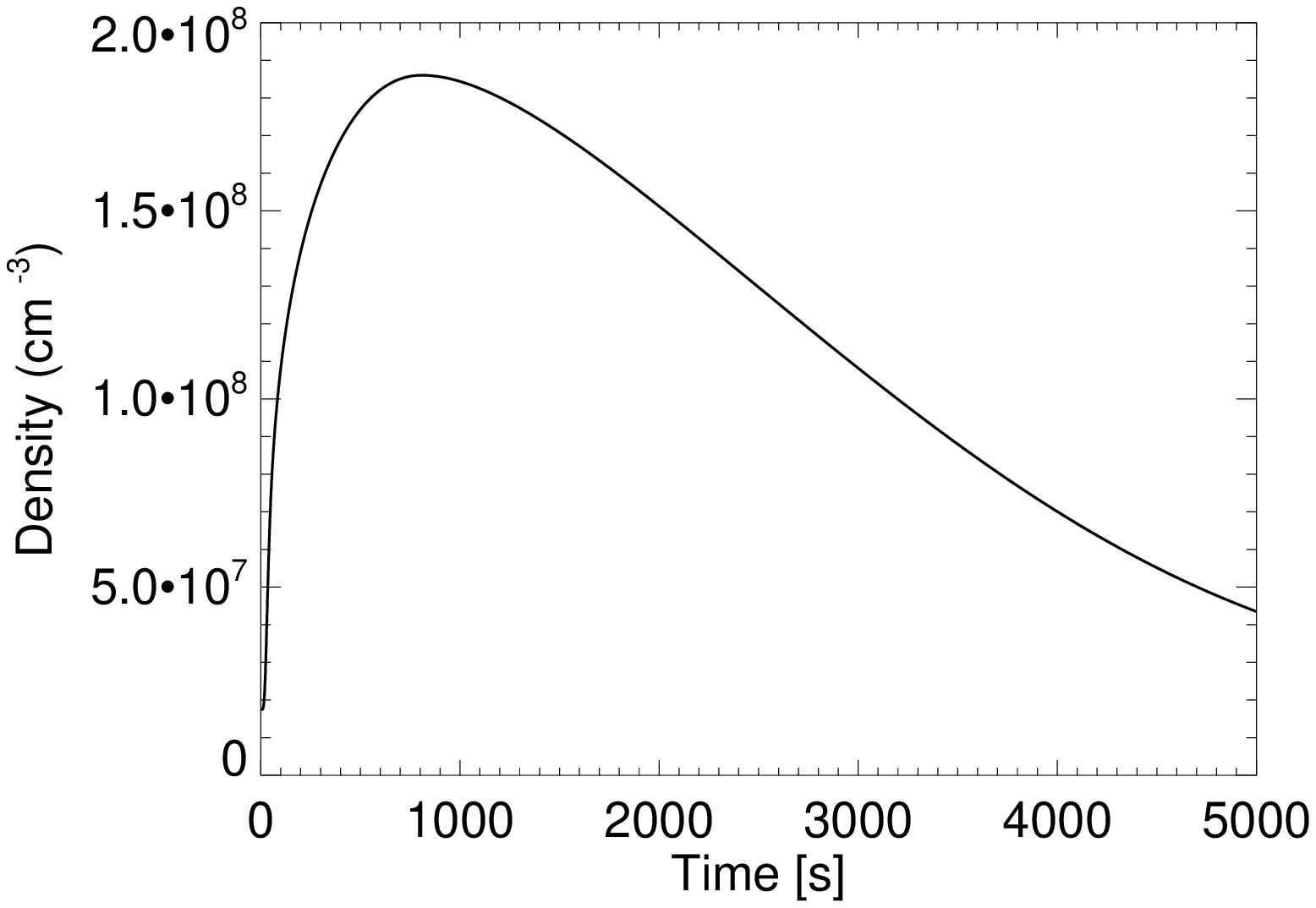}}
 \subfigure[]{\includegraphics[width=8cm]{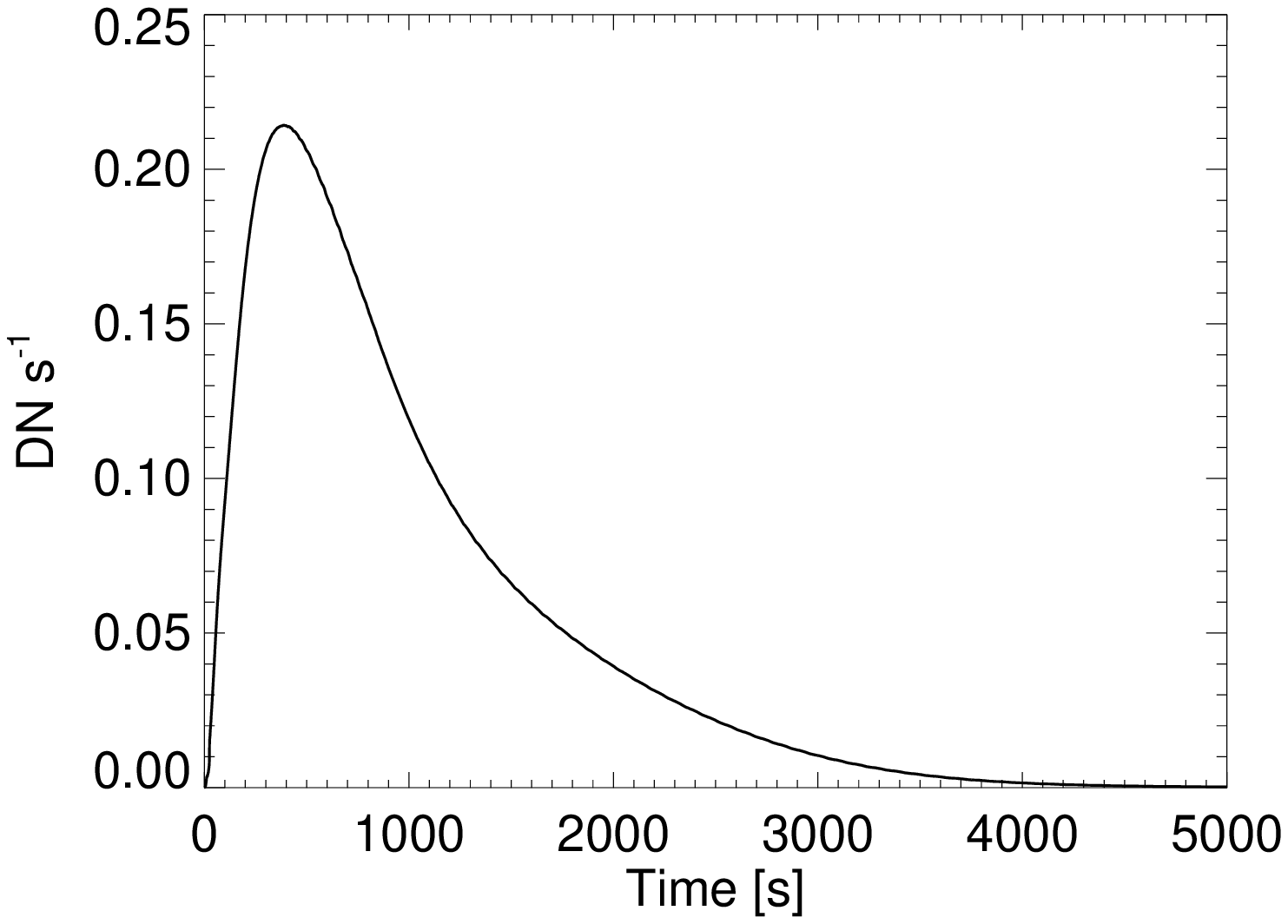}}
 
 \caption{\small Time evolution of (a) temperature, (b) density and (c) AIA 94 \AA\ intensity
of a single strand, heated by a pulse of 0.003 erg cm$^{-3}$ s$^{-1}$, and 50 s
duration. }
 
\label{fig3}
\end{figure} 
   
To this purpose, we use EBTEL (Enthalpy Based Thermal Evolution of Loops), a zero dimensional time-dependent hydrodynamic model (\citealt{Kli08}, \citealt{car12}). The model has no spatial resolution, and describes the evolution of the average physical properties of the plasma confined in a single coronal flux tube. The model assumes that the loop is symmetric with respect to the apex, and therefore describes half of it.
A physical key concept in the model is the enthalpy which has the main role in \aftr{transferring} energy budget into and out of the corona. The enthalpy will certainly not produce or eliminate energy. 
Variations in heating rate will affect transferring of mass between chromosphere and corona.  Any changes in heating rate will cause an increase or decrease in the heat flux and any excess or deficit in downward heat flux related to the transition region radiation loss will consequently derive an up flow enthalpy flux or will be compensated by a downward enthalpy flow respectively.
The energy equation is:
\begin{equation}
\frac{\partial E}{\partial t}=\frac{\partial (Ev)}{\partial s}-\frac{\partial (Pv)}{\partial s}-\frac{\partial F}{\partial s}+Q-n^{2}\Lambda(T)+\rho g_{\|}
\end{equation}
in which
\begin{equation}
E=\frac{3}{2} P+\frac{1}{2} \rho v^{2}
\end{equation}
E is the combination of thermal energy and kinetic energy, P is pressure, v is the bulk velocity, F is the heat flux, Q is the volumetric heating rate, n is the electron density, $ \Lambda $  is the radiation loss function and $g_{\|}$ is the gravity component which is along the magnetic field. \aftr{We assume classical conductivity, but we have ascertained that saturated conductivity does not change the results (and in particular the light curves, see Section~\ref{sec:res}) significantly on the relatively long time scales of our modeling.}
The evolution for density and pressure can be described fully versus the coronal parameters as we integrate the energy equation over the coronal part of the loop once and separately over the transition part. 
\begin{equation}
\frac{1}{\gamma-1} \frac{d\bar{P}}{d t}=\bar{Q}-\frac{1}{L}(R_{c}+R_{tr})
\end{equation}

\begin{equation}
\frac{1}{\gamma-1} \frac{d \bar{n}}{dt}=\frac{n_{0}v_{0}}{L}=-\frac{\gamma-1}{2KT_{0}L\gamma}(F_{0}+R_{tr})
\end{equation}
where \={P} and \={n}  are the averages of pressure and density, $R_{c}$ and $R_{tr}$  are the radiation losses in corona and transition region respectively, F is the heat flux v is the bulk velocity and 0 indexes describe the values in the base of the corona.
This set will be completed if we insert an equation of state.
\begin{equation}
\frac{1}{\bar{T}} \frac{d\bar{T}}{d t}=\frac{1}{\bar{P}} \frac{d\bar{P}}{dt}-\frac{1}{\bar{n}} \frac{d\bar{n}}{dt}
\end{equation}

We use EBTEL to model the evolution of the plasma confined in single loop strands under the effect of short heat pulses. Figure~\ref{fig3} shows the evolution of average coronal temperature and density inside a single strand subject to impulsive heating. The strand is initially cool and tenuous, with a temperature of 0.24 MK and a density of $\sim 10^7$ cm$^{-3}$. A very small amount of heating ($10^{-6} \ erg\, cm^{-3} \,s^{-1}$) is constantly provided to keep the strand in equilibrium. In addition to this, we impose a heat pulse of triangular time profile having a duration of $\tau=50 \, s$  and an intensity peak of $h=0.003 \ erg\, cm^{-3} \,s^{-1}$.  This heating rate corresponds to a temperature of $\sim 2.5$ MK at the equilibrium according to the scaling laws of \citealt{Ros78}. The thermodynamic decay time according to \citealt{Ser91} and \citealt{Rea14} is $\tau \sim800 \,s$. We follow the evolution over $10^4 \,s$ , i.e. more than 10 decay times in this case.

In figure~\ref{fig3} the temperature rises abruptly to $\sim 3$ MK  as a consequence of the intense nanoflare.  The strong heat flux drives massive evaporation from the chromosphere to the corona and the strand begins to fill with plasma to a maximum density of $\sim 2 \times 10^8 \,cm^{-3}$. This is much lower than the equilibrium density of $\sim 3 \times 10^9 \,cm^{-3}$, because of the short duration of the heat pulse. The temperature declines as the nanoflare shuts off by the effect of both the radiation and plasma thermal conduction toward the cool chromosphere.
The density peaks later \aftr{(by $\sim 10$ min)} than the temperature because the evaporation continues for some time \citep{Rea14}. The strands finally enter a long phase of draining as the radiation loss - the cooling mechanism- overcomes gradually the thermal conduction \citep{car12}. The density decay is quite slower than the temperature decay. 

From EBTEL results we can derive the emission in the EUV. Figure~\ref{fig3} shows the light curve in the SDO/AIA 94~\AA\ channel. The light curve has a shape in between the evolution of the temperature and of the density, because the emission is a function of both of them. \aftr{So, the peak of the emission occurs $\sim 5$ min later than that of the temperature.} This is a light curve of a single strand.  
When we look at the light curve of a pixel, we are summing the light curves of many strands that are intercepted along the line of sight in that pixel.
 
A single pixel may contain tens to several hundreds of strands each heated impulsively. We assume that the distribution of the heat pulses is described by a power law \citep[e.g.,][]{Hud91}:
\begin{equation}
dN = E^{-\alpha}dE
\end{equation}
where $dN$ is the number of events per energy interval $(E, E+dE)$ and $ \alpha $ is the power law index.
Each strand is ignited independently of the others, therefore at random times and at random intensities, according to the power law frequency distribution. Since we are unable to constrain the times and intensities, our approach is to generate a large number of different realizations of the same light curve that consists of overlapping light curves of a given number of strands. Each light curve is related to a random pulse extracted from the power law distribution and has a random start time. We generate groups of realizations, one for a given power law index, pulse duration and number of strands.

As mentioned in Section~\ref{sec:obs}, the total loop length is fixed to $5 \times 10^{9} \,cm$.
We choose two possible values of the power law index, i.e. $\alpha=1.5$ and $\alpha=2.5$.
We choose either of two possible pulse durations, i.e. a short ($\tau=50$ s) one and a long ($\tau=500$ s) one, with respect to typical plasma cooling times. The other key parameter is the number of strands that are heated along the line of sight, and in particular we consider three possible values, N=10, 100 and 1000. 
We assume that each strand is heated only once during our total time lapse. In this view, the number of strands is also the number of heat pulses. \aftr{There is discussion about the frequency of the heat pulses inside a single strand, whether the repetition time is large (low frequency) or small (high frequency) with respect to the typical cooling times \citep{Kli15}. Our scenario is basically low frequency. However, we address only the high temperature emission, and similar results might be obtained with more frequent pulses, with a delay not much longer than 1000 s (see Fig.\ref{fig3}c). This might be in agreement with recent constraints \citep[e.g.,][]{Car14,Car15}.}

The observed light curves fluctuate around a steady value. When we generate events at random times and we overlap them, we have to wait that a steady state is reached. We checked that a safe time lapse to reach steady state is half an hour ($1800 \,s$). Therefore we assume that all the events occur at random times within the total observation time, i.e $3576 \,s$ plus half an hour, therefore in a time range $0<t<5376 \,s$.
When we compare our results with the observations we will consider only the final $3576 \,s$, in which we are sure that the simulated emission is as steady as the observed one. In the following all the model light curves will start from $t= 1800 \,s$ that will be assumed as the reference time.  

\begin{figure*}[!t]
\centering

\setlength\fboxsep{0pt}
\setlength\fboxrule{2pt}
\includegraphics[width=16cm, height=10cm]{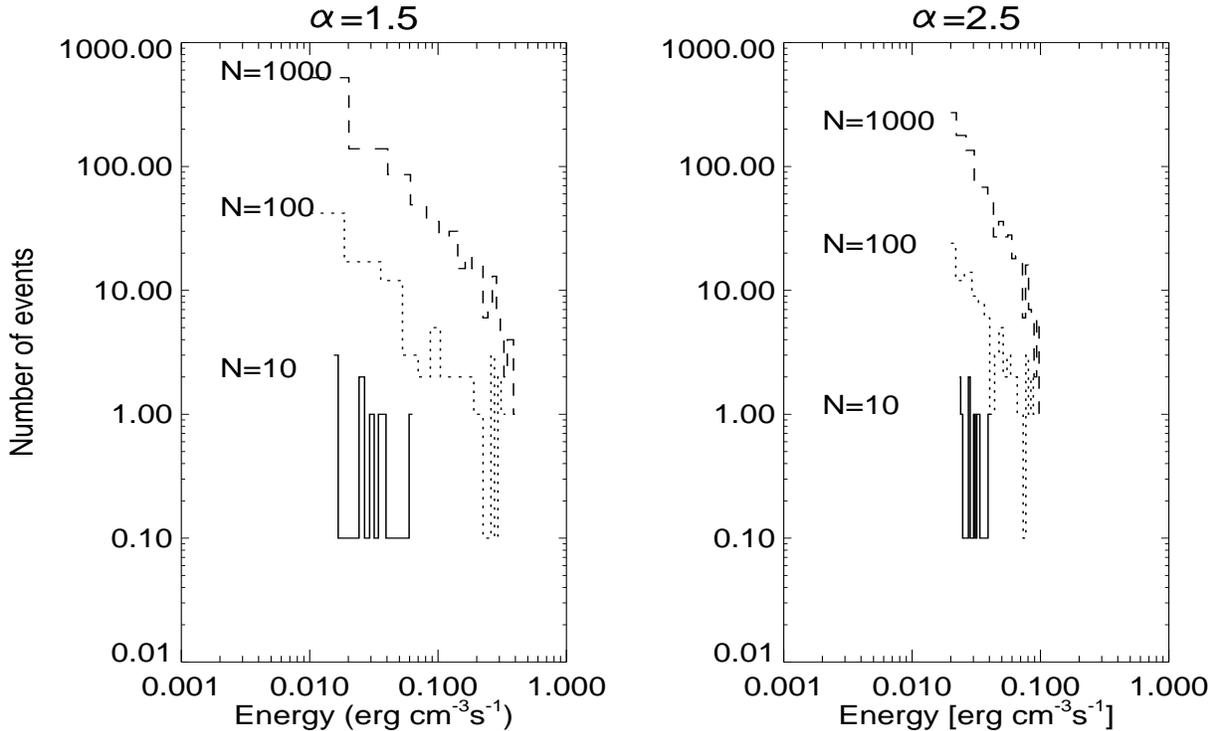}
\caption{Left to right: The distribution of simulated energies for power law index $\alpha=1.5$, $\alpha=2.5$.
For both the cases the duration of the heat pulses is $\tau=50 \,s$. } 

\label{fig4} 
\end{figure*}

We tune the range and height of the power law distributions so as to produce a bundle of strands that has the average temperature of the observed active region loops, i.e. about 3 MK. Figure~\ref{fig4} shows the energy rate distributions of the pulses obtained for different numbers of strands and for the shorter pulse durations. 
Instead of producing one model for each energy rate, i.e. for each realisation, we have preferred to generate a grid of EBTEL models, choosing the parameters so as to span reasonably all possible loop conditions in the energy range.
We then use a binary search to find the closest value of each energy from the original power law distribution to the one in the grid.

For each of the 12 combinations of $ \alpha$, $\tau$,  and $N$, a package of 10000 pairs of light curves (two for each realization) is produced. We compare each pair of light curves to the observed ones and choose the best matching ones with the methods described in the following.


\subsection{Probabilistic Neural Network} \label{bozomath}
 
\begin{figure}[t]
\plotone{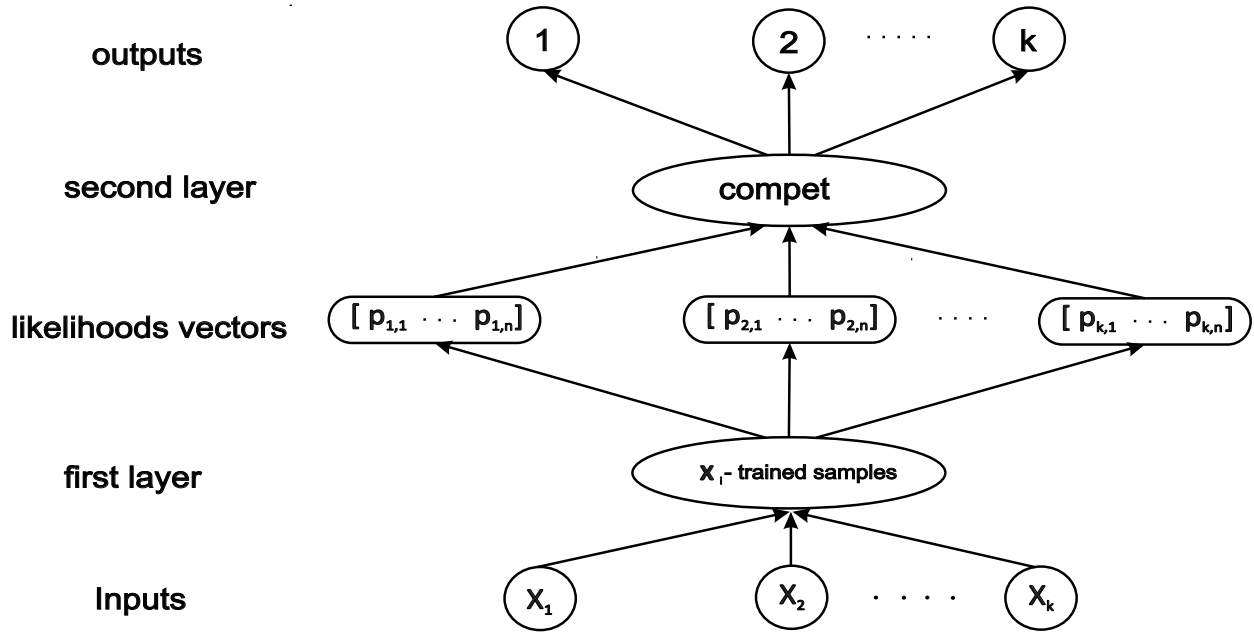}
\caption{An Architecture of Probabilistic Neural Network} 
\label{fig5}
\end{figure}

We employed PNN (Probability Neural Network), a kind of Artificial Network which is suitable in classifying and identifying the samples.
Recently it has been successfully applied in every field of science as a classifier machine. 
The main and the first step in performing a comparison by the tool is to train the network by training samples. During the training session the network will learn the possible determined classes. 
The architecture of a probabilistic Neural Network that shows a complete training session is shown in figure~\ref{fig5}.

When an input vector is fed to the network, the first layer computes the distance from the input vector to the training samples. This produces a vector which indicates how close the input is to the training samples. 
The second layer sums the contribution for each class of inputs and produces its net output as a vector of probabilities. Finally, the Compete transfer function on the output of the second layer picks the maximum of these probabilities and produces 1 (positive identification) for that class and  0 (negative identification) for non-targeted classes. 
In this way the Network will correspond each tested sample to \aftr{its} own class of trained samples (\citealt{Baz08}, \citealt{Taj12}).

\aftr{The Probabilistic Neural Network is a supervised algorithm which needs to be trained before being used for classification. Therefore, the greater the number of training samples we provide to feed the network, the more accurate the output of PNN will be. Its performance is based on estimating the probability density function from sample patterns, which implicitly consists of calculating the distances between an input vector with the other training samples. So, it seems to somehow do just like a nearest neighbor classifier \citep{Mon92} while comparing to other kind of classification methods. For this reason, it can better deal with irrelevant features.}

\subsection{Cross Correlation} \label{sec:cross}

The other method that we use is cross correlation which is simply a way to measure how similar the signals are. This kind of approach has been recently applied to the analysis of coronal observations (\citealt{Via13}). 
Two input vectors of x and y may be cross-correlated as a function of time lag L as:\textbf{•}

\begin{equation}
corr_{x,y}=\frac{\sum^{M-\vert{L}\vert-1}_{k=0}(x_{k+\vert{L}\vert}-\textbf{\={x}})(y_{k}-\textbf{\={y}})}{\sqrt{\sum^{M-1}_{k=0}(x_{k}-\textbf{\={x}})^{2}\sum^{M-1}_{k=0}(y_{k}-\textbf{\={y}})^{2}}}   \quad \textrm{for} L < {0}
\end{equation}

\begin{equation}
corr_{x,y}=\frac{\sum^{M-\vert{L}\vert-1}_{k=0}(x_{k}-\textbf{\={x}})(y_{k+\vert{L}\vert}-\textbf{\={y}})}{\sqrt{\sum^{M-1}_{k=0}(x_{k}-\textbf{\={x}})^{2}\sum^{M-1}_{k=0}(y_{k}-\textbf{\={y}})^{2}}}   \quad \textrm{for} L \geq {0}
\end{equation}



where \={x} and  \={y} are the averages of the vectors $x=(x_{0},x_{1},..., x_{M-1})$ and $y=(y_{0},y_{1},..., y_{M-1}) $, respectively ($M$ is the number of data points).
 
The function $c\_correlate$ within the IDL software enables us to find the correlation values between two selected samples at any given time lag. The highest similarity between two samples will be indicated by the maximum cross correlation value.

\section{Results} \label{sec:res}

The next step is to compare these light curves to the observed ones and to find the best match for each combination, with both the comparison methods independently. Our key point is to find the realization that best matches the light curves in both channels at the same time.


With the PNN method we are not able to compare couples of light curves simultaneously, but only one model light curve with one observed light curve at a time. Since we want to match simultaneously light curves in two different channels, our solution has been to join each light curve in one channel to the corresponding light curve in the other channel. \aftr{So, we first normalize each light curve to its maximum value, and then we stitch the end point of one to the first point of the other.} This is done for both the simulated and observed light curves.

Afterwards, we trained the network with the simulated light curves as training samples. The observed light curves are then fed the network as the tested samples to classify with respect to the trained samples.
The process of classifying the data is done for the light curve of the single and of the   row of pixels separately.

\begin{figure}[!ht]               
\centering
 \subfigure[]{\includegraphics[width=8cm]{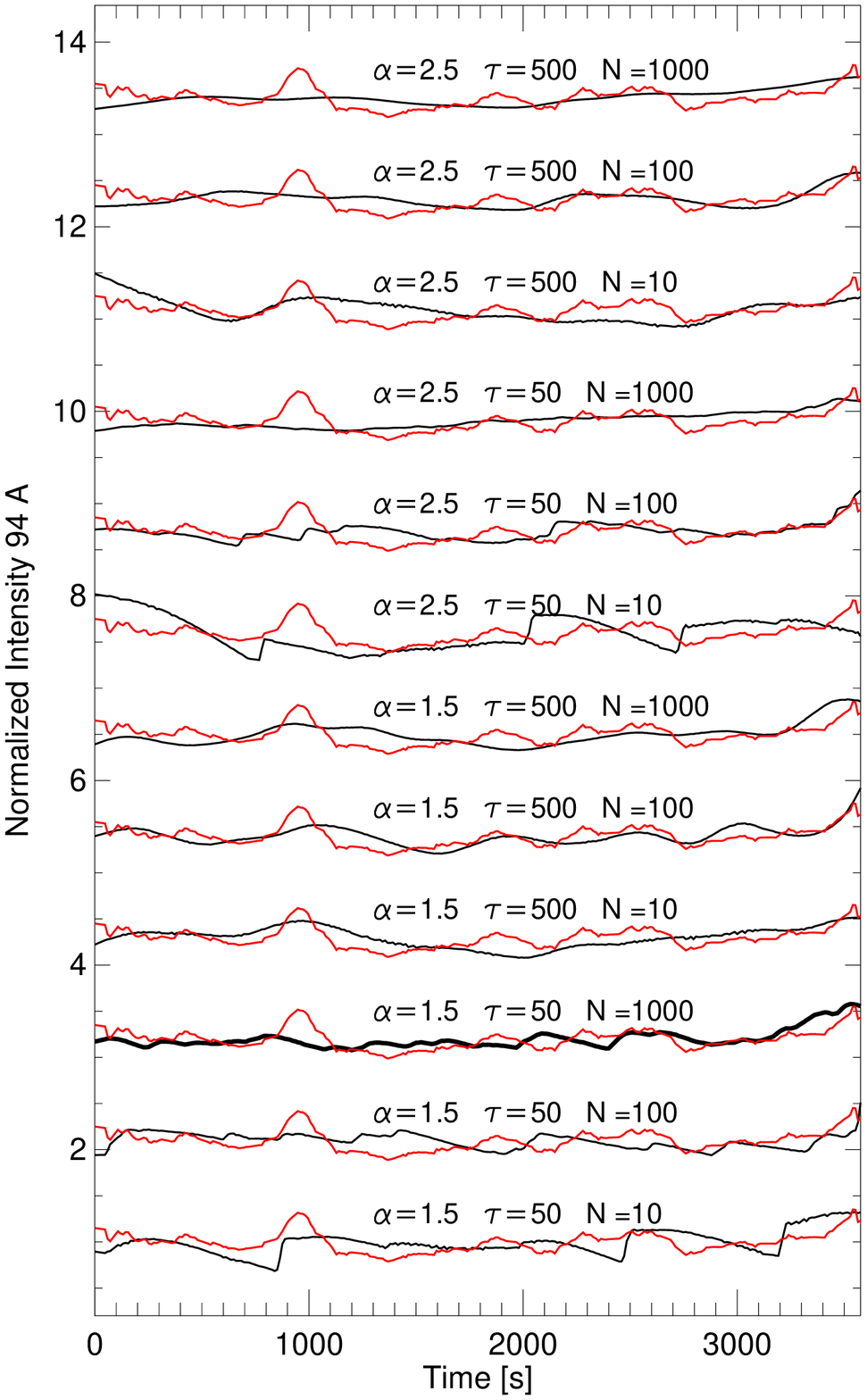}}
 \subfigure[]{\includegraphics[width=8cm]{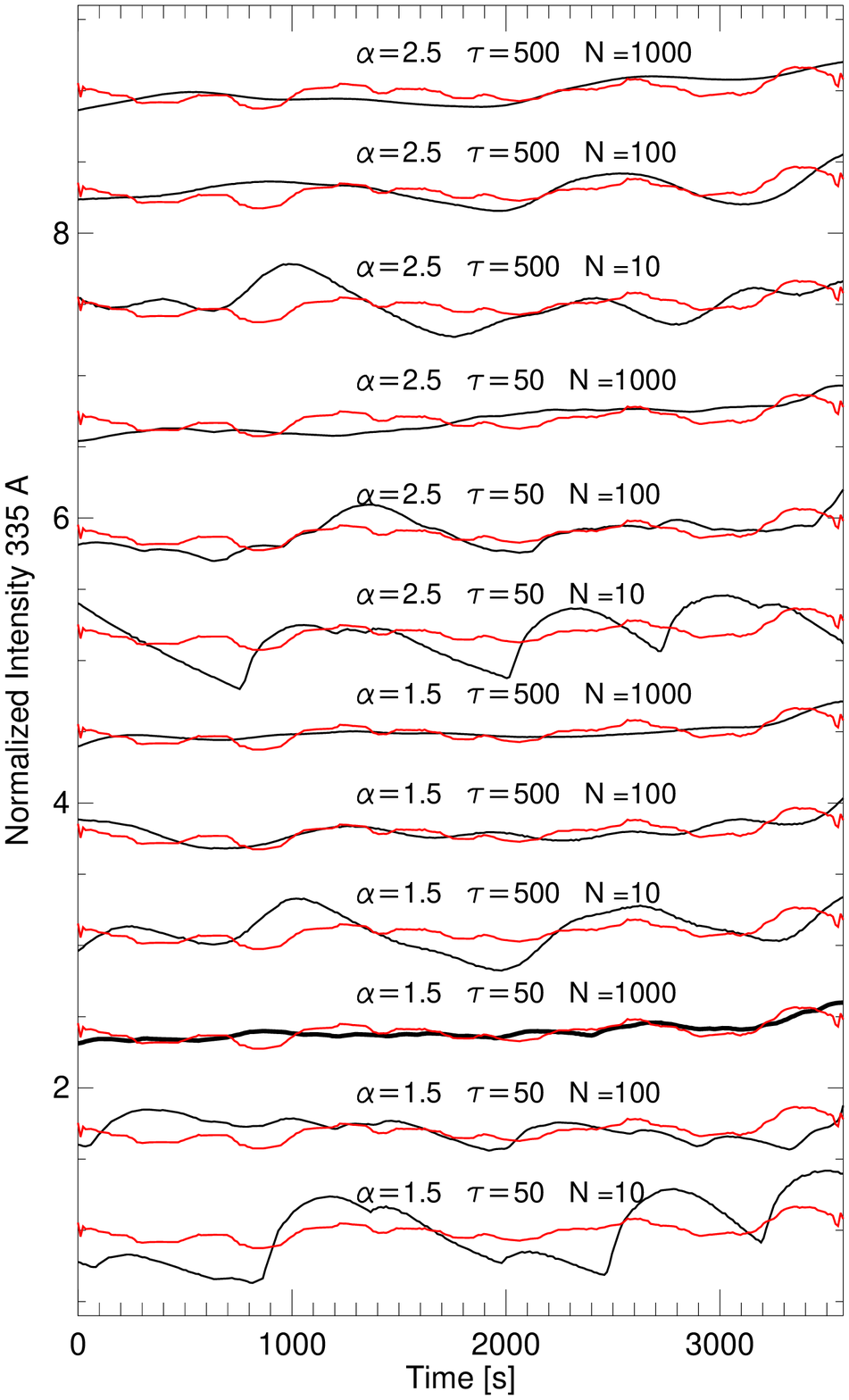}} 
 \caption{\small \aftr{{\it Black solid lines:} Light curves from realisations for each set of parameters ($\alpha, \tau$ and N) that best match the observed ones in the single pixel (red lines) according to the PNN method. The light curves in both the 94 \AA\ (left column) and 335 \AA\ (right) channels are shown.  Comparison of observed (red lines) light curves to the best model ones (black lines) found by the network for each set of parameters ($\alpha, \tau$ and $N$).
For a better visual comparison, the intensities are normalized to the average and shifted each by a different value, and the observed light curves have been smoothed with a boxcar of 8 points. The best absolute match is marked (thick black lines). }} 
\label{fig6}
\end{figure}

The output of the network finds the best choice among the available set of simulated patterns which best resemble the data and labels it with its own corresponding key parameters. \aftr{In Fig.~\ref{fig6}, for each of the 12 sets of parameters, we show the light curves of the realisation that best matches the observed ones according to the PNN method.  For a better visual comparison, we have applied normalization, smoothing and shifting procedures, but we remark that the PNN compares the realizations with the original light curves (after a normalization only).  }

In general, the PNN is unable to find simulated patterns that perfectly match the observed ones, not even in one channel. The PNN chooses the best solution as the one that shows the best match of the overall general patterns.
\aftr{As mentioned above, we let the method find the best solution for each set of parameters. It remains to be found the best absolute solution. We might rank the best solution on its overall ability to reproduce the details of the observed features, and in particular the amplitude, shape and time scale of the observed bumps in both channels. From a visual inspection of Fig.~\ref{fig6} we realize that the best solutions are not equivalent. Those with the long pulse duration, with a small number of strands and with a steep distribution all show too broad features, which do not fit the observed features on the smallest time scales, especially in the 94 \AA\ channel. Also the solutions with the steeper distribution ($\alpha=2.5$) are in general unable to reproduce the variability on short time scale. A small number of strands determines too strong bumps in the 335 \AA\ channel.  The best absolute solutions appear those with the flatter energy distribution, and shorter pulse duration. Among these, the one with the largest number of strands yields the lowest total root mean square deviation (RMSD) from the observed light curves defined as:}


\begin{equation}
\mbox{RMSD} = \sqrt{ \sum_i^M \frac{(\overline{R}_i-\overline{O}_i)^2}{M}}
\end{equation}
\aftr{where $\overline{R}_i$ are the model intensities normalised to their average (for each channel), $\overline{O}_i$ are the observed intensities normalised to their average. 
So, eventually, the best set of parameters found with PNN is [$\alpha=1.5$, $\tau=500$ and $N=1000$] (thick black lines in Fig.~\ref{fig6}), for which we obtain $\mbox{RMSD} = 0.21$.}

Two points here are worthy of being pointed out. One is the fact that the network is not sensitive to denoising or smoothing \citep{Taj12}.  
Another point is that the output of the network does not change even when we reverse the order of stitching the light curves. This means that the network is robust in its performing.

\begin{figure}[!ht]               
\centering
 \subfigure[]{\includegraphics[width=8cm]{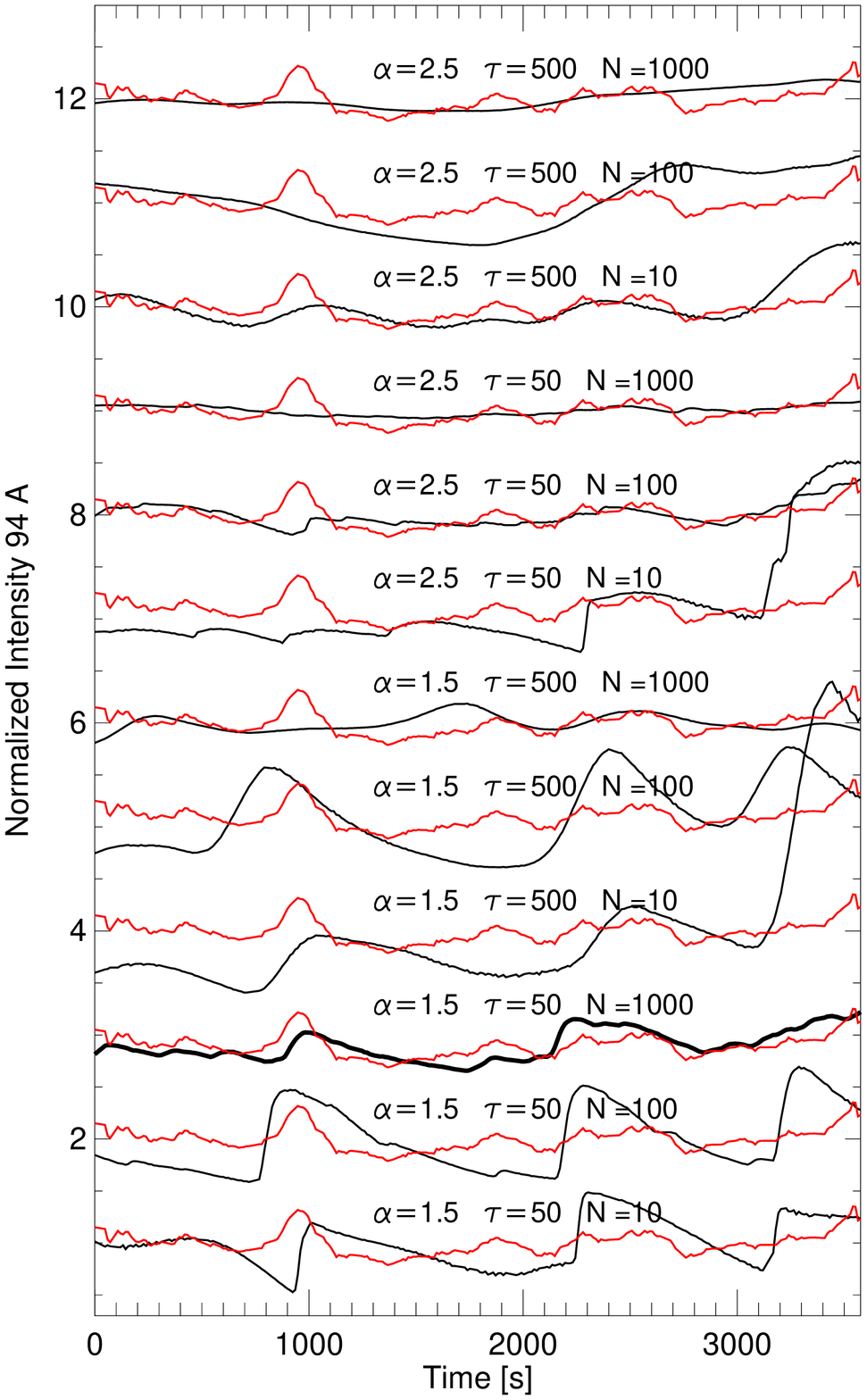}}
 \subfigure[]{\includegraphics[width=8cm]{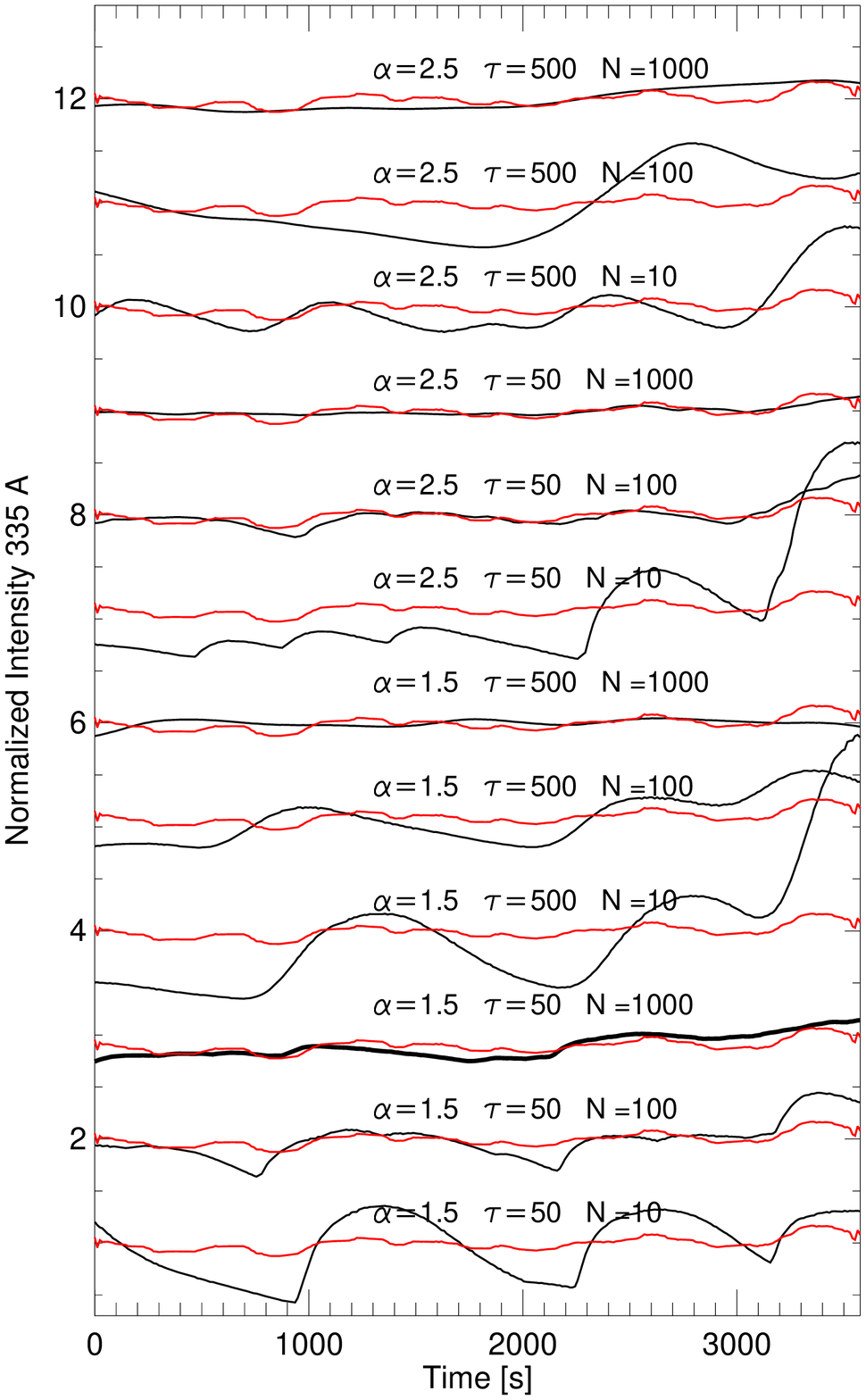}}

 \caption{ \aftr{As Fig.~\ref{fig6}, according to the cross-correlation method. }}
 
\label{fig8}
\end{figure}

We make an alternative comparison using the simple cross-correlation technique described in Section~\ref{sec:cross}. 
With this method we can compare the simulated and observed light curves of both channels simultaneously without joining them. In a given channel each model light curve is time-shifted and cross-correlated with the observed one (either the single pixel or the pixel row), and the cross-correlation value is computed. As we did for PNN, we do this for each set of parameters. The best match is given by the realization that provides the highest sum of cross-correlation values found for the two channels and the same time lag. \aftr{The best matching realisations for each set of parameters found with cross-correlation are shown in figure ~\ref{fig8} (see Fig.~\ref{fig6} for comparison). }


\aftr{The best absolute matching (and the highest cross correlation value) is obtained by the same set of parameters as that found with the PNN method, i.e., $\alpha=1.5$, $\tau=50$,  $N= 1000$, although with a different realisation (thick black lines in Fig.~\ref{fig8}). However, for this realisation we obtain $\mbox{RMSD} = 0.24$, which is slightly higher than that obtained with the PNN method. }

\aftr{We obtain very similar results when we compare the model realizations with the  observed light curves extracted from the row of pixels, i.e. the same best set of parameters with both PNN and cross-correlation. In the following, we will consider the best solution found for the single pixel and with the PNN method as the best absolute one. }

\begin{figure}[t]
\includegraphics[width=12cm]{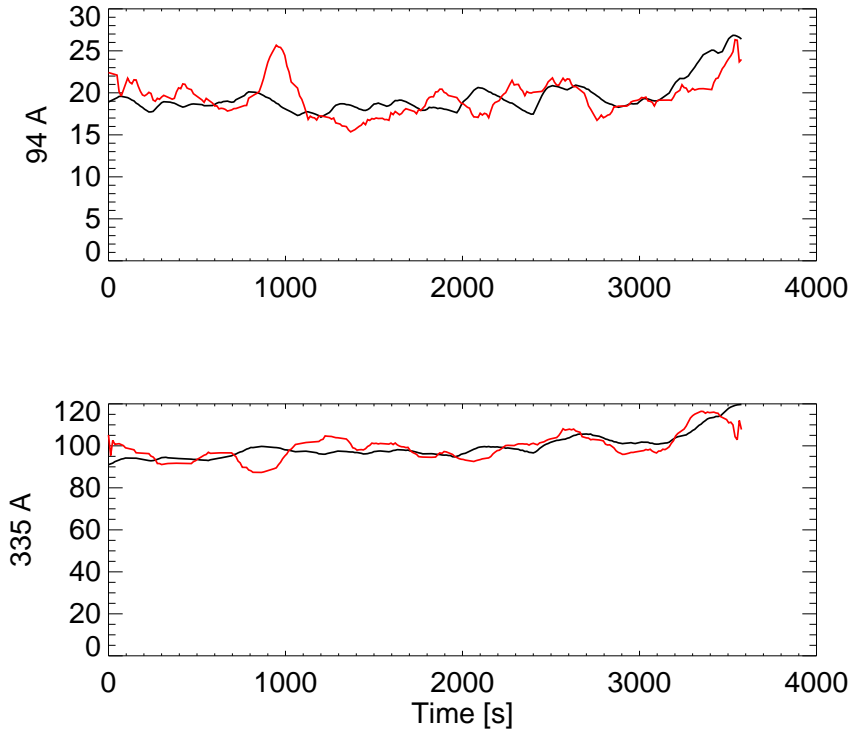}
\caption{\aftr{Model light curves (black lines) in AIA 94~\AA\ channel (upper panel) and 335~\AA\ channel (lower panel) for the case $\alpha=1.5$, $\tau=50$ and $N=1000$, best matching the observed ones (red lines) found with the PNN method. 
This is the best absolute match. The model intensities are scaled to match the average observed intensities, by assuming a total cross-section area of 0.56 and 0.52 pixels, respectively.}} 
\label{fig7}
\end{figure}

\aftr{It is interesting to make considerations about the absolute intensity values. We first compare the ratio of the mean observed intensities with those of the best realisation. We obtain $(I_{335}/I_{94})_{obs} \approx 5.1$ vs  $(I_{335}/I_{94})_{mod} \approx 5.5$. The agreement is remarkable, the percent difference ($\sim 7$\%) being less than the average fluctuations of the 94 \AA\ light curve ($\sim 15$\%). The slightly higher emission observed in the 335 \AA\ channel might be simply due to some diffuse emission along the line of sight. Fig.~\ref{fig7} shows the best matching model alone. In this figure we go back to the original intensities with no normalisation. To report the model results to the observations we have to make an assumption about the cross section of the strands. We find that we need a cross section of 0.56 and 0.52 pixels in the 94 \AA\ and 335 \AA\ channels, respectively, to match the best model to the observed light curves. In the assumption of 1000 equal and independent strands, this is equivalent to find that each strand has thickness of $\sim 10$ km. This becomes a lower limit if the strands are not entirely independent, i.e., if the same strand is heated more times during our time lapse (see Section~\ref{sec:ebtel}).} 
We should also keep in mind that we have a logarithmic spacing in our sampling of the number of strands, and therefore this value of the thickness should be taken with care.  

\section{Discussion and conclusions}

In this work we analyse the time evolution of the EUV emission in the core of an active region, which shows evidence for a very hot ($T > 5$ MK) plasma component \citep{Rea11}. This hot component might be a signature of the occurrence of rapid but intense heating releases, which bring the plasma to such high temperature for short times. In that active region we consider the light curves at the maximum time resolution in three SDO/AIA channels picked up either in a single pixel or in a row of pixels where the emission evolves coherently. 

We try to match the observed light curves with the emission derived from specific loop modelling. 
The simultaneous presence of very hot plasma and steady emission indicates that we might have storms of events with a broad range of energy distribution.
In the light of this evidence, our choice has been to describe the evolution in a scenario of loops made by bundles of independent strands each heated for a time shorter than the typical plasma cooling times (e.g., \citealt{Gua10}). We assume that each strand is tenuous and cool at the beginning and is heated only once, at a random time, by a heat pulse of random intensity. It is then left free to decay. This is equivalent to fewer strands heated repeatedly but \aftr{not at high frequency, i.e., after time intervals longer than the cooling and draining times \citep[][see also Section~\ref{sec:ebtel}]{War10,Kli15}.}

Since we address the time evolution only, and no spatial issues, we preferred to consider the very efficient approach of 0D loop modelling, that describes the evolution of the average quantities of the coronal plasma contained in a loop magnetic flux tube. The output of the model is the evolution of the average density and temperature, that we use to derive the light curves in relevant channels to be compared with the observed ones. An important issue is the choice of the free parameters. We assume that all strands have the same length, which is constrained from the observation. We observe mostly straight bright structures in the core of the active region deep in the disk, so we assume semicircular strands that stand vertically from the surface.  The other important parameters are the intensity of the heat pulses, their duration and the number of strands. In the framework of randomly occurring events, we assume that the heat pulses are distributed as power laws. We assume two possible values of the power law index, i.e. a shallower ($\alpha =1.5$) and a steeper ($\alpha = 2.5$) one.
We then normalize the intensity and range of the distribution so to have an average heating rate that is able to produce a loop plasma at 3 MK on average, according to the loop scaling laws. We set two possible duration of the heat pulses, a short (50 s) and a long (500 s) one. The number of strands changes logarithmically, from a few (10), to a relatively large number (1000). For each of the two pulse distributions with different $\alpha$ and of the two pulse durations, we generate a grid of 0D models. For each model we derive the light curves in the AIA 94~\AA\ and 335~\AA\ channels. The next step has been to choose a number of strands and to combine randomly the corresponding number of light curves in a channel, according to one of the pulse distributions and for one pulse duration. So we randomly pick up an intensity from the intensity distribution, and a random start time of the pulse, uniformly distributed in a time range of 10000 s. For each set of parameters we derive 10,000 different  realisations, i.e. random combinations of light curves in the 94~\AA\ and 335~\AA\ channels. Each couple of light curves has been compared to the couple of the observed ones. The comparison has been made independently with two different methods, one based on artificial intelligence, the other on a simple cross-correlation. 
We do not address a perfect match of the simulated and observed light curves, that would require much larger sets of realisations. We let the methods find the best realisations for each set of the parameters. Then, we compare 
these best cases and pick up the one that is able to reproduce patterns globally similar to the observed ones, and in particular variations with similar amplitude and timescales and similar shapes of the local emission bumps or dips.

\begin{figure}[!ht]               
\centering
 \subfigure[]{\includegraphics[width=8cm]{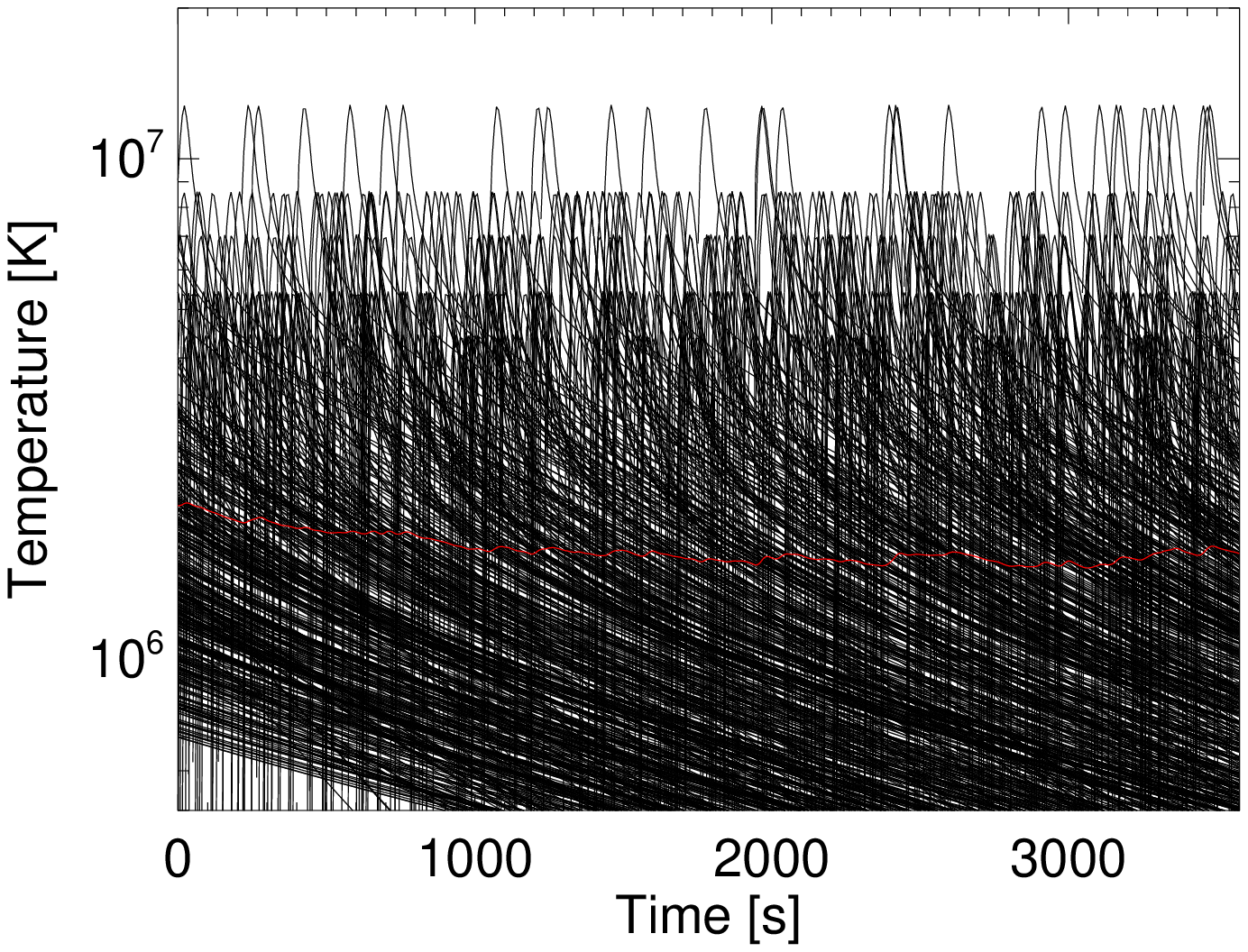}}
 \subfigure[]{\includegraphics[width=8cm]{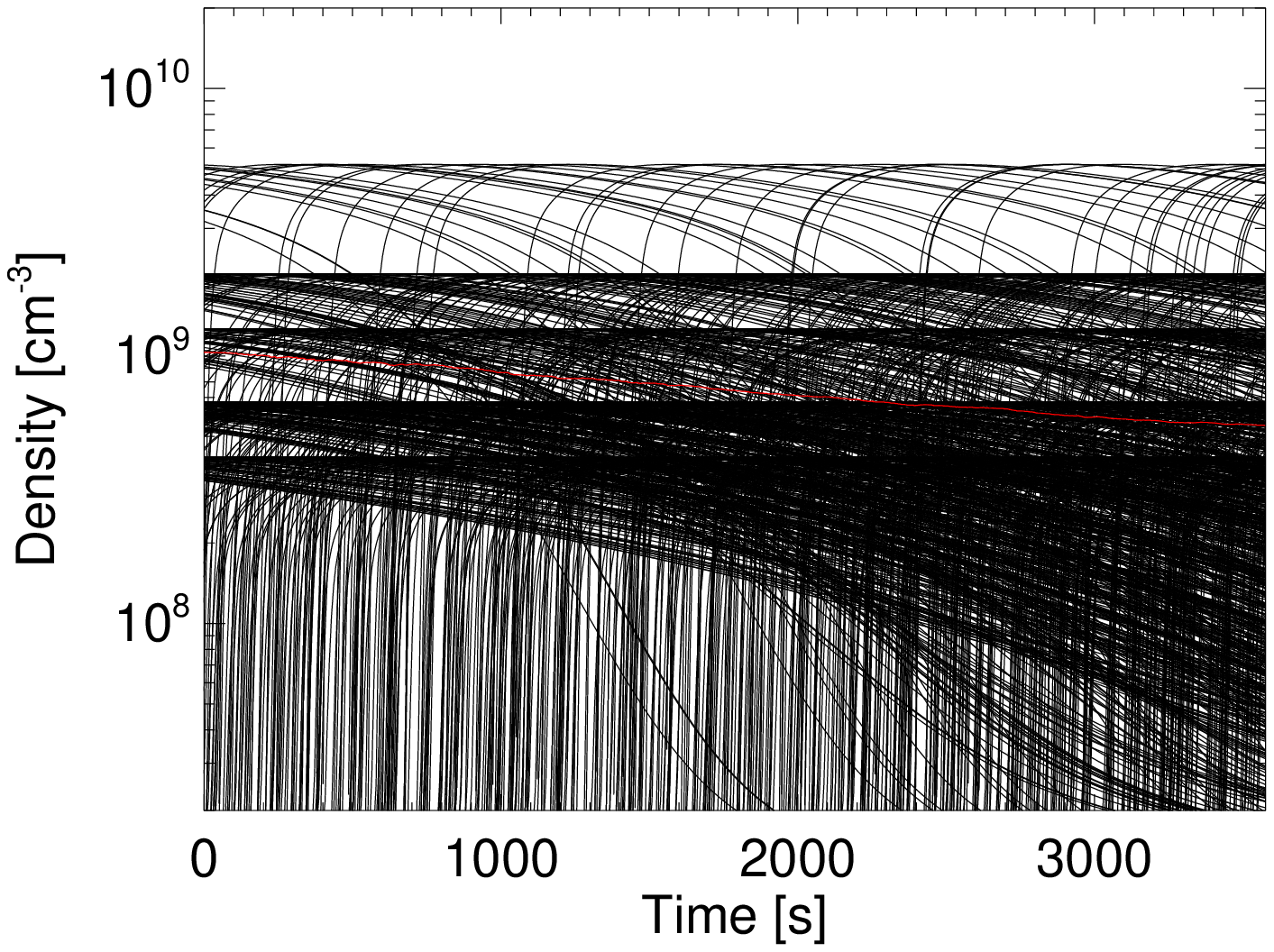}}
 \subfigure[]{\includegraphics[width=8cm]{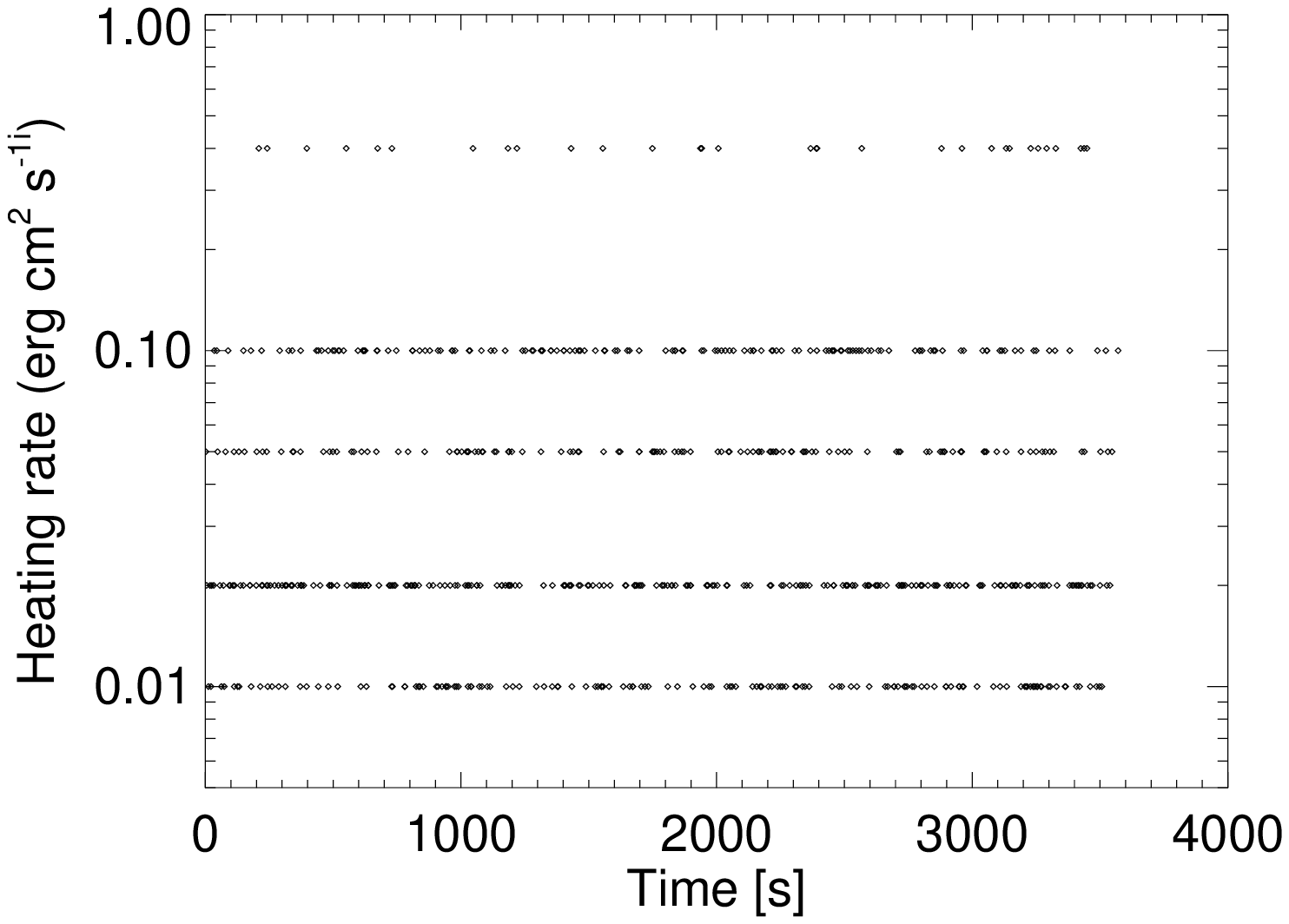}}
 \caption{\small  Time evolution of (a) temperature, (b) density and (c) heat pulses of all the randomly heated strands that overlap to build the model light curve in Fig.~\ref{fig7}.  The red lines are the average values. Time has been shifted by 1800 s.}
\label{fig9}
\end{figure} 


The parameter set of the realisation that best matches \aftr{qualitatively and quantitatively the observed light curves} is a shallow power law index ($\alpha =1.5$), a short pulse duration (50 s) and relatively large number of strands (1000). This realisation has been singled out with the PNN method, minimizes the deviations from the observational data, is able to reproduce well many features of the observed light curves, and delivers a ratio of the intensities that is consistent with the observed ones.
Realisations with the same set of parameters best match the light curves both of the single pixel and of those averaged over a row of a few pixels.

The distribution of events is able to describe the presence of both many weak and few strong events, that explain both the rather smooth light curves and the presence of a small amount of hot plasma at the same time. 
The relatively small duration of the heating release, of the order of 1 minute, is in agreement with recent finding from observations and modelling (\citealt{Tes13}, \citealt{Tes14}). A relatively high number of heated strands is preferred, and is able to reproduce well the rather steady emission. \aftr{We find that the intensities from this combination of parameters are compatible with strands $\sim 10$ km thick or more, to be compared with recent measurements ($\sim 100$ km) from high resolution observations ~\citep{Bro13}.}

\begin{figure}[!ht]               
\centering
 
 \subfigure[]{\includegraphics[width=8cm]{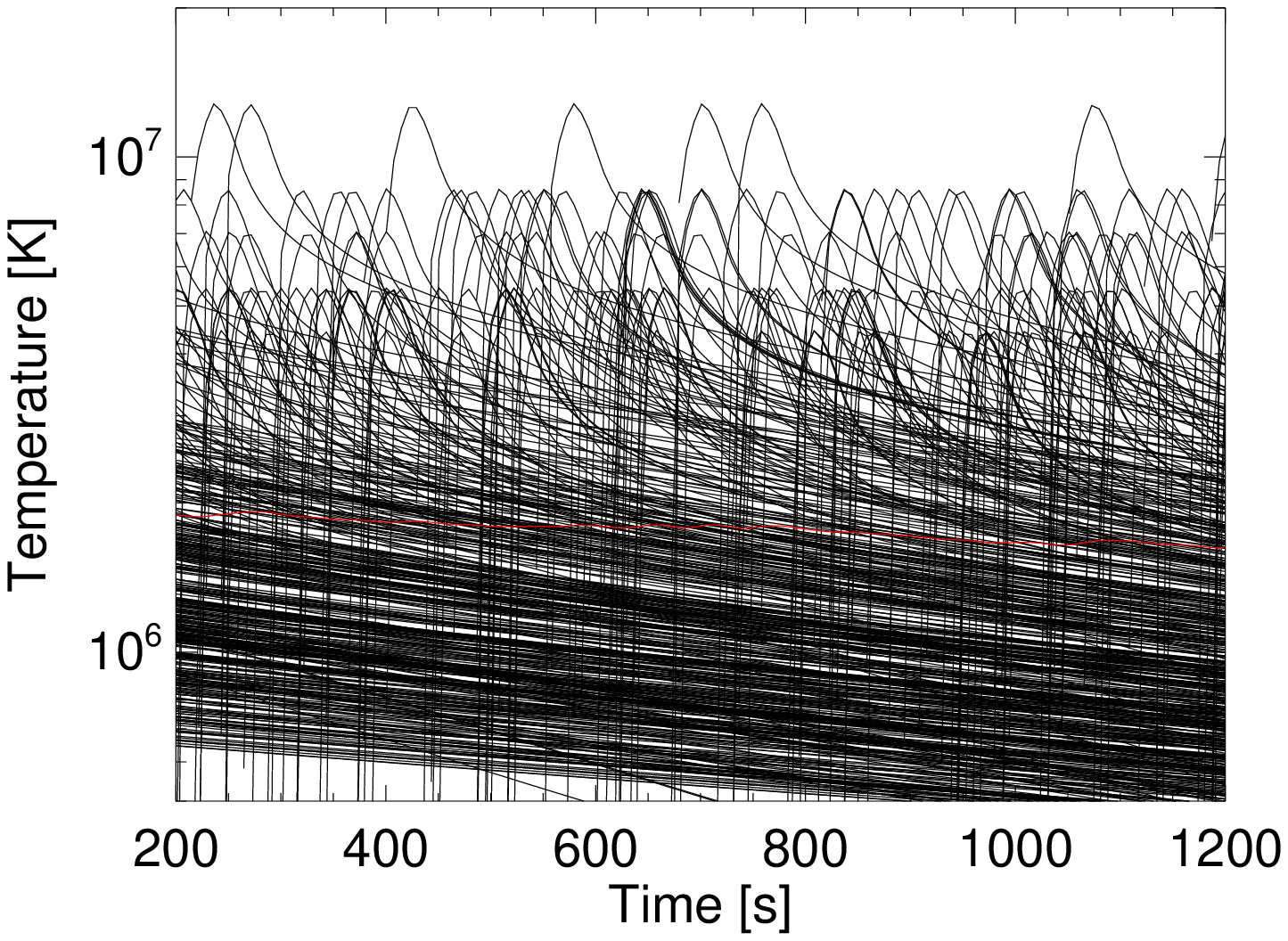}}
 
 \subfigure[]{\includegraphics[width=8cm]{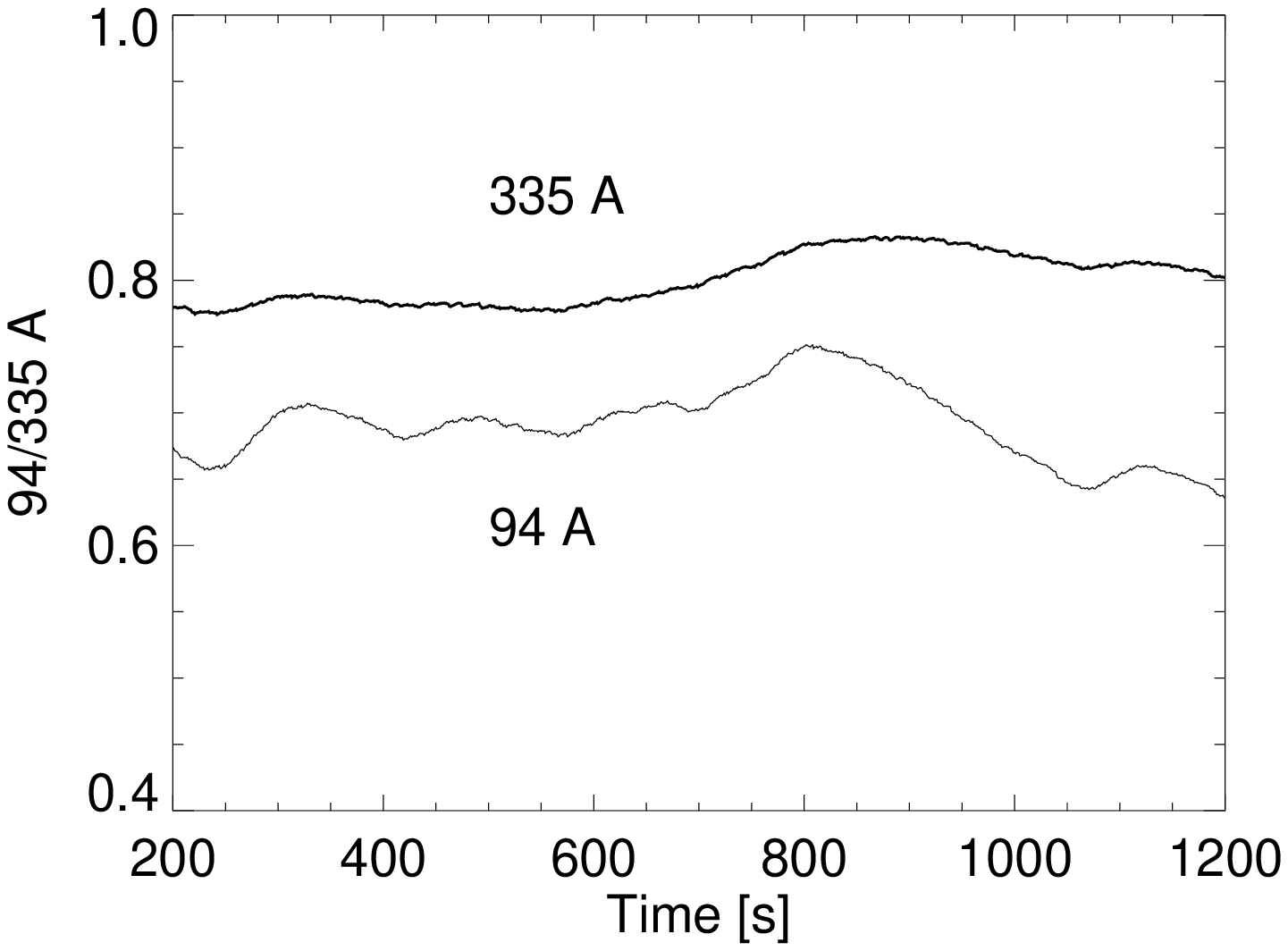}}
 \subfigure[]{\includegraphics[width=8cm]{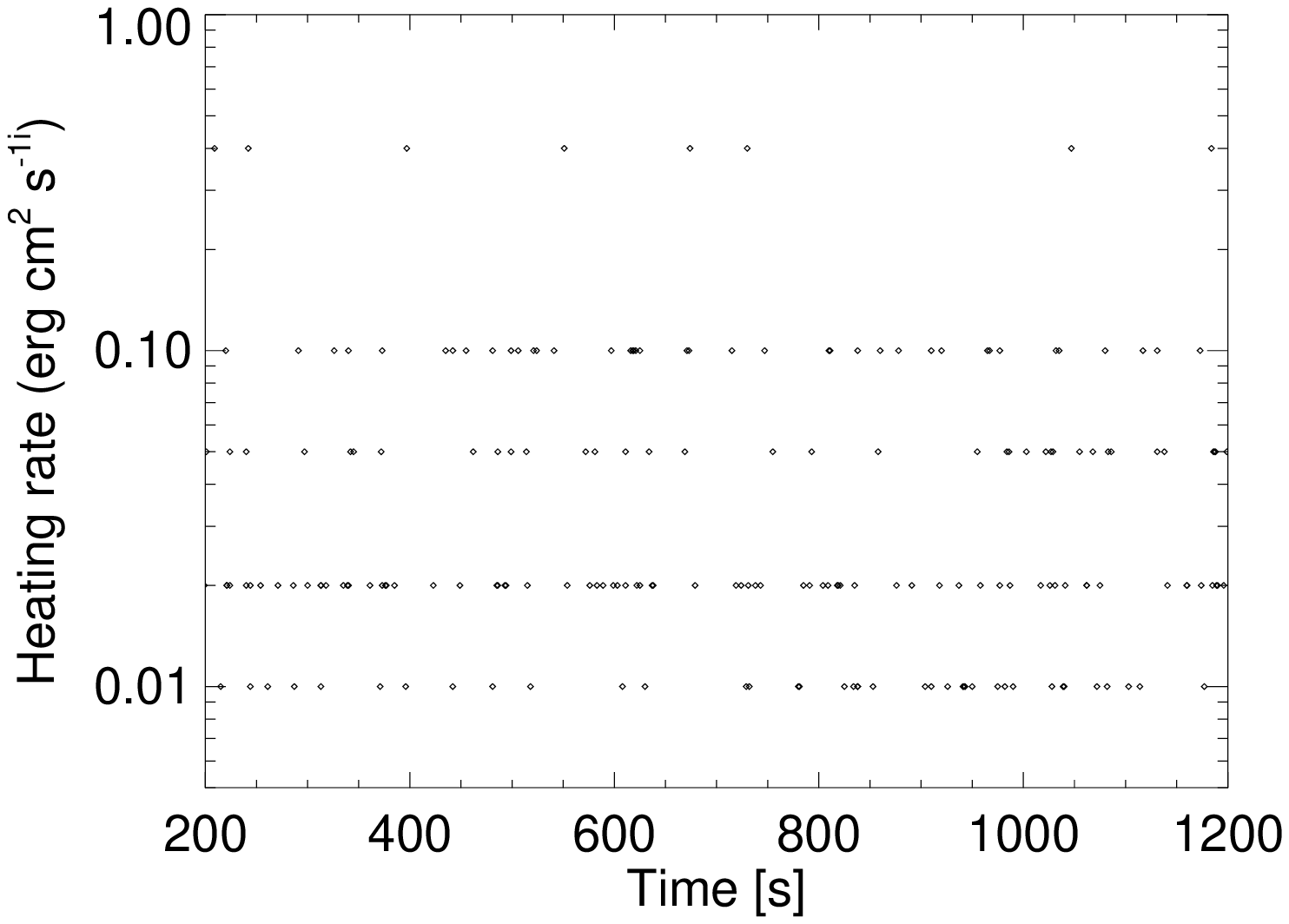}}
 \subfigure[]{\includegraphics[width=8cm]{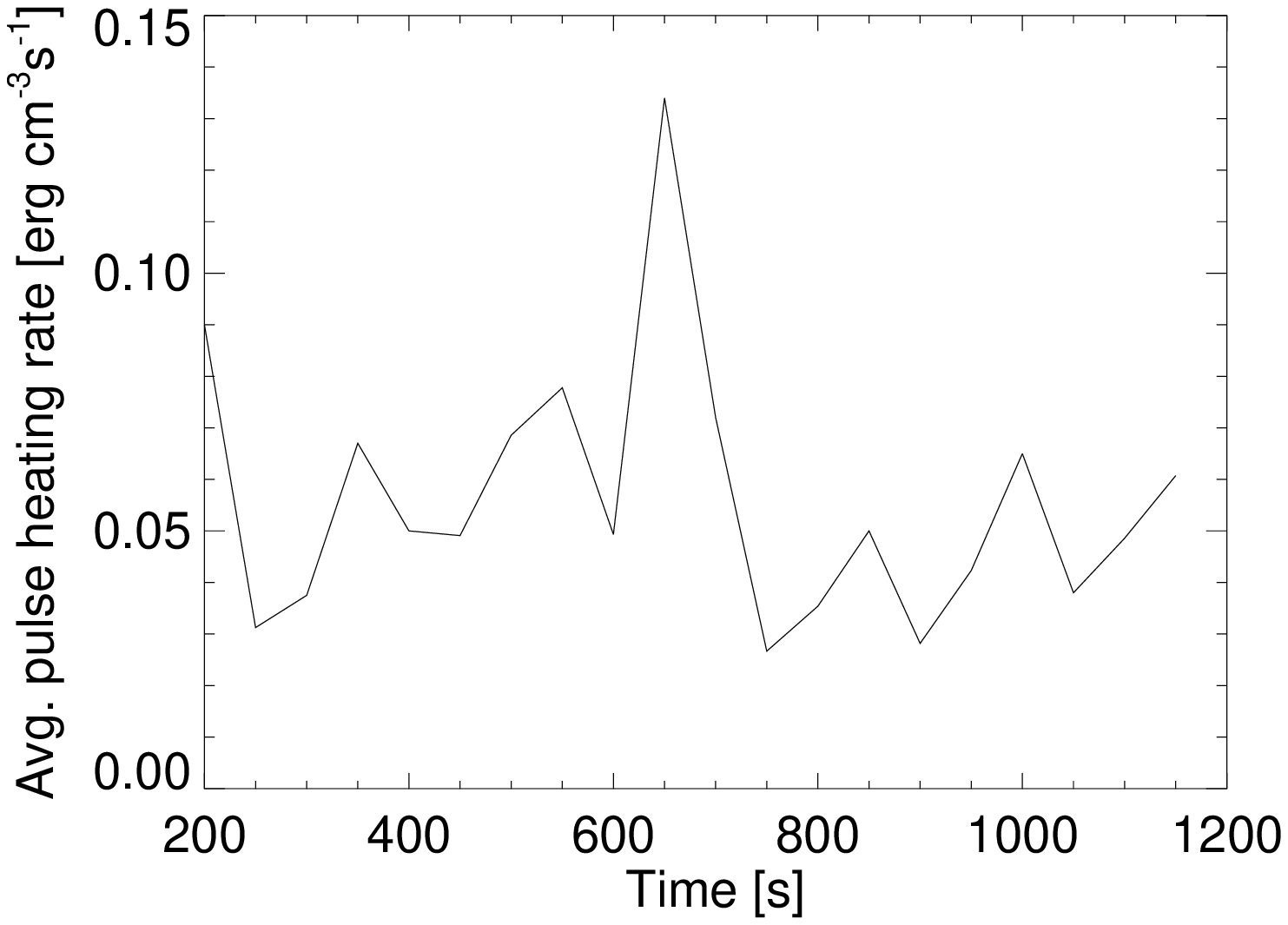}}
 \caption{\small  Enlargement of Fig.~\ref{fig9} in a time range of 1000 s, showing (a) temperature, (b) intensity of 94~\AA, and 335~\AA\ (see Fig.~\ref{fig7}),  (c) the heat pulses and (d) the average heat rate of each pulse over bins of 50 s. }
\label{fig10}
\end{figure}

The simulated light curves that best match the observed ones show that the emission fluctuates more in the 94~\AA\ channel, and it is smoother in the 335~\AA\ channel, which is more sensitive to cooler plasma. However, we can see a large scale similarity in the global trends. Figure~\ref{fig9} shows the evolution of the plasma and heating event properties for all the modeled strands. In the temperature plot,  among the multitude of lower temperature events, we clearly distinguish a smaller number of events that bring the temperature above 10 MK for short times. They are \aftr{consistent} with the detection of a small and filamented amount of very hot plasma in this region \citep{Rea11}. It is interesting to search for signatures of physical processes in the light curves. Figure~\ref{fig10} zooms in a 1000 s time range of Fig.~\ref{fig9} and shows the temperature events, the distribution of heat pulses and the evolution of the average event heating rate, with a 50 s time binning. While we do not see any obvious correspondence between the light curves and the first two quantities, we clearly see a correlation of the event heating rate with the trends observed in the light curves. In particular, we see a train of heat bumps that anticipates a train of emission bumps by $\sim 200$ s in the 94 \AA\ channel. \aftr{This time lag is of the same order as the delay between the temperature peak and the emission peak for a single strand shown in Fig.~\ref{fig3}a,c (see also Section~\ref{sec:ebtel}), ultimately due to the more gradual evolution of the density.}
We might therefore infer that strong fluctuations in the 94~\AA\ channel probably mark a previous increment of heating episodes with a delay of a few minutes. This signatures is  present also in the 335~\AA\ channel, but  is much less significant.

Overall, the analysis presented here shows results that are \aftr{consistent} with previous works. The short and infrequent heat pulses are largely consistent with the presence of cooling plasma for most of the time, which was detected in Hinode/XRT observations \citep{Ter10} and in SDO/AIA observations \citep{Via12}. In the former work a good match with observations is obtained with Monte Carlo simulations including most energetic random pulses at average time distance of 360 s from each other. From Fig.~\ref{fig10} we can count a number of $\sim 30$ highest pulses (0.4 erg cm$^{-3}$ s$^{-1}$, with comparable peak temperature and duration with those in \citealt{Ter10}) in $\sim 1200$ s, corresponding to an average time distance of $\sim 150$ s. This higher frequency is not included in \cite{Ter10} and might be consistent for the core of an intense active region. The model also involves the presence of very hot plasma as detected in the same region analysed here in SDO/AIA observations \citep{Rea11}. Our analysis obtains additional information about the number, distribution and intensity of the heating events and about the fine loop structuring, and figures out possible signature of the heating directly detectable in the light curves. 


This works improves also on the previous analysis with artificial intelligence methods \citep{Taj12}, because we use specifically a loop model as basic model,  because we address matching simultaneously the light curves in two different channels, and because we cross-check with a different comparison method, namely cross-correlation.

A further improvement on our analysis could be the attempt to include also some spatial information, i.e. the coherence of the signals in the same loop structures. This requires using more detailed loop models that describe the confined plasma with spatial resolution.

\acknowledgements{We thank the referee for very constructive comments, and Dr. T. Di
Salvo for suggestions. E.T., F.R.,  A.P. acknowledge support from  italian Ministero
dell'Universit\`a e Ricerca.P.T. was supported by contract SP02H1701R from Lockheed-Martin to the Smithsonian Astrophysical Observatory, and by NASA grant NNX15AF50G. SDO data supplied courtesy of the SDO/AIA consortia. SDO is the first
mission to be launched for NASA's Living With a Star (LWS) Program.}


\bibliographystyle{aa}
\bibliography{biblio}

\end{document}